\documentclass[%
twocolumn,
showpacs,
 amsmath,amssymb,
 aps,
 prb,
floatfix,
 citeautoscript,
]{revtex4-1}

\usepackage[dvipdfmx]{graphicx}% Include figure files
\usepackage[dvipdfmx]{color}
\usepackage{dcolumn}% Align table columns on decimal point
\usepackage{bm}% bold math
\usepackage{braket}
\begin{document}

%\preprint{APS/123-QED}

\title{Density matrix renormalization group study of the $\nu=1/3$ edge states in fractional quantum Hall systems}
%\thanks{A footnote to the article title}%

\author{Takuya Ito}
 \email{ito@cmpt.phys.tohoku.ac.jp}%Lines break automatically or can be forced with \\
\author{Naokazu Shibata}%
\affiliation{%
 Department of Physics, Tohoku University, Sendai 980-8578, Japan
 }%

%\collaboration{MUSO Collaboration}%\noaffiliation

\date{\today}% It is always \today, today,
             %  but any date may be explicitly specified

\begin{abstract}
The edge states in the fractional quantum Hall systems at filling factor $\nu=1/3$ are studied by the density matrix renormalization group method. It is shown that the density oscillation induced by the local boundary condition at the edge is characterized by the wave number of the minimum magnetoroton excitation, and this structure is partially reconstructed with the change in the confinement potential shape.
In particular, the $\nu=1$ counterpropagating edge channel appears with the change
in the chemical potential, which is consistent with recent experiments on heat transport. 
The stability of the bulk states against the change in the number of electrons confirms that 
the bulk part of the fractional quantum Hall state is incompressible, while the edge state is compressible.

%begin{description}
%\item[Usage]
%Secondary publications and information retrieval purposes.
%end{description}
%\pacs{73.43.-f, 05.10.Cc}
\doi{10.1103/PhysRevB.103.115107}

\end{abstract}

                             % Classification Scheme.
%\keywords{Suggested keywords}%Use showkeys class option if keyword
                              %display desired
\maketitle

%\tableofcontents

\section{Introduction}

Since the discovery of the fractional quantum Hall (FQH) effect\cite{Tsui1982,Laughlin1983},  many interesting properties of two-dimensional (2D) electrons have been reported in quantum Hall systems.\cite{Sarma1997} Extensive studies have shown that the bulk part of the FQH state has an excitation gap, while gapless excitations exist along the edge of the 2D system. These unique low-lying excitations are theoretically described by the one-dimensional (1D) model called the chiral Luttinger liquid\cite{Wen1992}, and the transport properties of the FQH states are expected to be determined by the excitations of the edge state\cite{Chang2003}. Recent theoretical and experimental works, however, have shown the results are not simply explained by the above conventional edge picture. The noise measurement of the edge current of the $\nu=1/3$ FQH state has reported the presence of neutral heat transport\cite{Inoue2014} which was originally predicted in hole conjugate states\cite{MacDonald1990,Johnson1991}, and the bulk heat transport was also reported even in the FQH state\cite{Altimiras2012,Inoue2014}. In addition, the tunnel current experiment has shown the sample-dependent exponent of the $I\mathchar`-V$ power law, which is not consistent with the theoretical predictions\cite{Chang1996,Grayson1998,Hilke2001,Chang2001}. Besides these reports, exact diagonalization studies have indicated the edge reconstructions by the competition between the repulsive Coulomb interaction and confinement potential\cite{De1994,Wan2002,Wan2003,Joglekar2003,Wei2020}, suggesting the formation of extra counterpropagating edge modes which modify the transport property\cite{Kane1994,Kane1995,Bid2010}. 
To understand these results quantitatively, systematic study of the edge states is needed. 

The FQH effect is realized in a strong magnetic field where the kinetic energy of electrons is quenched into highly degenerate Landau levels. The resulting macroscopic degeneracy 
leads to the failure of analytical perturbation theory, and numerical analysis has been used to solve many-body problems caused by the Coulomb interaction. 
To systematically analyze the edge state, however, we need to deal with a large system beyond the limitation of exact diagonalizations since the length scale of the density oscillation induced by the confinement potential is much larger than the system size available for exact diagonalizations and it is difficult to realize the bulk part of the FQH state 
between the two counterflowing edge channels.

In this paper we use the density matrix renormalization group (DMRG) method\cite{White1992,White1993} to calculate the ground state wave function of large systems with more than 70 electrons under various confinement potentials and clarify the stability of the edge state and the conditions for the edge reconstruction. Although the DMRG method was originally designed for 1D interacting systems, it has been successfully applied to FQH states of two-dimensional systems under strong magnetic fields\cite{shibata2003,shibata2003a,Feiguin2008,Hu2012}. We calculate the ground state of the two-dimensional electron gas in torus geometry at filling factor $\nu=1/3$ and show that the reconstruction of the edge state occurs in good agreement with previous exact diagonalization studies, and the bulk part of the FQH state is stable under a shift of the chemical potential within a certain range, which is consistent with the experimentally observed Hall conductivity plateaus of the FQH state. 

\section{Model and method}

\begin{figure}[t]
\centering
\includegraphics[scale=0.22]{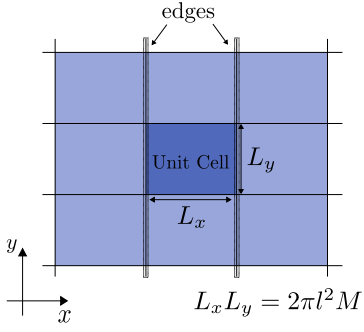}
\caption{Two-dimensional electron system on torus geometry used in the calculation. The central blue region indicates the unit cell. $L_x$ and $L_y$ are the lengths of the unit cell along the $x$ and $y$ axes, respectively. $M$ represents the number of magnetic flux quanta in each unit cell. Replacing $\delta'_{i,j}$ with the usual Kronecker delta $\delta_{i,j}$ in Eq. (4) results in breaking the translational symmetry for the $x$ direction, and the edge states appears along the $y$ axis.}
\label{figure1}
\end{figure}

The system used for our calculation is illustrated in Fig.~1. The lengths of the unit cell along the $x$ and $y$ axes are given by $L_x$ and $L_y$, respectively. Enclosed magnetic flux quanta are represented by $M$, which is related to the system size through the relation $L_x L_y = 2\pi l^2 M$, with $l$ being the magnetic length.
To investigate the fundamental FQH edge state, we consider the lowest Landau Level (LLL) and assume the electron spin is fully polarized. After the projection onto the LLL, the one-particle wave function in the Landau gauge is written as 
\begin{align}
\varphi_j({\bm r}) &=\left [\frac{1}{L_y\sqrt{\pi}l}\right ]^{1/2} \times \notag \\ & \sum_{n \in \mathbb{Z}}  {\rm exp} \left[ i\frac{(X_j+nL_x)y}{l^2} -\frac{(x-X_j-nL_x)^2}{2l^2} \right ],
\end{align}
where $X_j$, defined as $X_j = \frac{2\pi l^2 j}{L_y}$, is the center coordinate of the $x$ direction for each one-particle wave function.
The Hamiltonian of a two-dimensional electron gas in a high magnetic field is described only by the Coulomb interaction as 
\begin{align}
H &= \frac{1}{2}\int d{\bm r}_1 \int d{\bm r}_2 V({\bm r}_1-{\bm r}_2):\hat{\rho}({\bm r}_1)\hat{\rho}({\bm r}_2): \notag  \\ &= \frac{1}{2}\sum_{j_1,j_2,j_3,j_4} \int _0 ^{L_x} dx_1 \int _0 ^{L_x} dx_2 \int _0 ^{L_y} dy_1 \int _0 ^{L_y} dy_2 \times \notag \\ &\varphi^{*}_{j_1}({\bm r}_1)\varphi^{*}_{j_2}({\bm r}_2)V({\bm r}_1-{\bm r}_2)\varphi_{j_3}({\bm r}_2)\varphi_{j_4}({\bm r}_1)c^{\dagger}_{j_1}c^{\dagger}_{j_2}c_{j_3}c_{j_4} \notag \\ &=\frac{1}{2}\sum_{j_1,j_2,j_3,j_4}A_{j_1,j_2,j_3,j_4}c^{\dagger}_{j_1}c^{\dagger}_{j_2}c_{j_3}c_{j_4},
\end{align}
where the Coulomb interaction $V({\bm r})$ and the 
coefficient $A_{j_1,j_2,j_3,j_4}$ are given as 
\begin{align}
V({\bm r}) = \sum_{n_x,n_y \in \mathbb{Z}} \frac{e^2}{\varepsilon}\frac{1}{|{\bm r}+n_xL_x{\bm e}_x+n_yL_y{\bm e}_y|},
\end{align}
 and
\begin{align} \label{wavefunc}
&A_{j_1,j_2,j_3,j_4} =  \delta ' _{j_1+j_2,j_3+j_4} \frac{1}{L_xL_y} \times \notag \\ &\sum_{{\bm q} \neq {\bm 0}} \delta'_{j_1-j_4,q_yL_y/2\pi}   \frac{2\pi e^2}{\varepsilon |{\bm q}|}  {\rm exp}\left[-\frac{q^2l^2}{2}-i(j_1-j_3)\frac{q_xL_x}{M}\right ],
\end{align}
respectively. ${\bm q}$ in Eq.(4) is the discrete wave vector, 
and $\delta'_{i,j}$ is the extended Kronecker delta, 
which is 1 if and only if $i = j+nM\ (n \in \mathbb{Z})$\cite{Yoshioka1984}. 

To introduce the edge on the torus system, we change the term $\delta'_{i,j}$ to the usual Kronecker delta $\delta_{i,j}$. This change removes the nondiagonal matrix element over the unit cell and breaks the translational symmetry of the $x$ direction, which results in the creation of edges on both sides of the system along the $y$ axis (see Fig.~1). Since changing from $\delta'_{i,j}$ to $\delta_{i,j}$ is interpreted as introducing a cut along the $y$ axis in the bulk, the obtained density oscillations 
on both sides of system are recognized as the edge states caused by
the boundary conditions.
We then extend the unit cell and analyse the effect of the confinement potrential shape in the next section. 

To deal with the large unit cell, we apply the DMRG method. This method enables us to iteratively expand the unit cell and obtain the ground state wave function within a desired accuracy that is determined by the number of remaining state $m$ in the calculation. We keep up to at least 200 basis states whose corresponding truncation error is in the range of $O(10^{-3})\mathchar`-O(10^{-5})$. 
We also checked the accuracy of our results by comparing ground state energy obtained by the exact diagonalizations
 up to system of $M=20$. 
 In the following sections, we take $e^2/\varepsilon l$ as units of energy.

\section{Results}

\subsection{Effect of the local boundary condition}

\begin{figure}[t]
\centering
\includegraphics[scale=0.65]{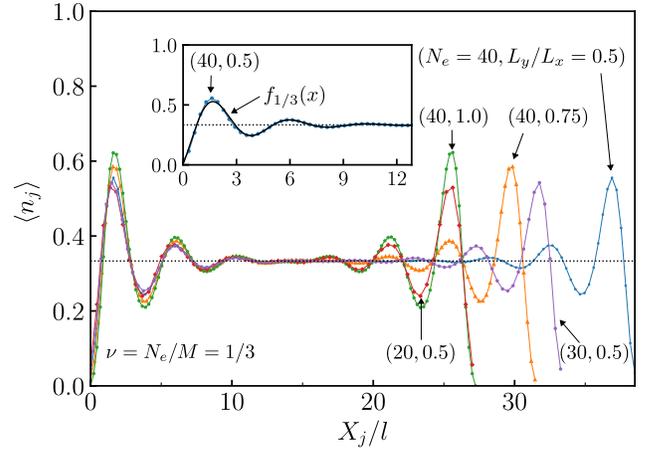}
\caption{Density oscillations of the $\nu=1/3$ FQH state induced by the boundary. $N_e=M/3$ is the total electron number in the unit cell, and $L_y/L_x$ is its aspect ratio. The black dotted line represents $\langle n_j \rangle=1/3$.
The thick black line in the inset shows the fitting function of Eq. (\ref{fitting}) for the result of $N_e=40,L_y/L_x=0.5$. }
\label{figure2}
\end{figure}

\begin{figure}[t]
\centering
\includegraphics[scale=0.65]{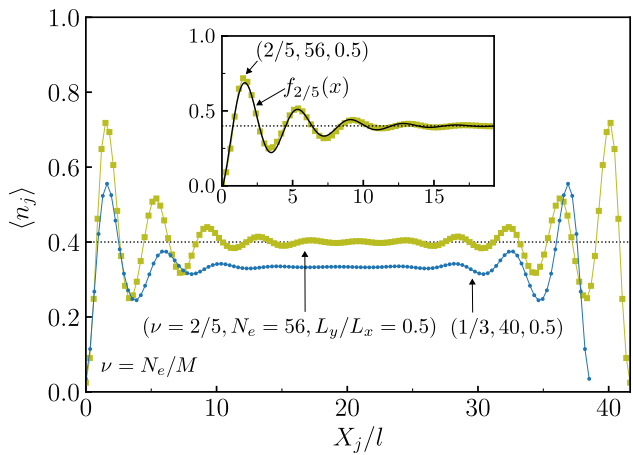}
\caption{Density oscillations of the $\nu=2/5$ FQH state 
induced by the boundary. $N_e=56$, and $L_y/L_x=0.5$. 
The black dotted line represents $\langle n_j \rangle=2/5$.
The blue line represents the result at $\nu=1/3, N_e=40, L_y/L_x=0.5$ for comparison. 
The thick black line in the inset shows the fitting function of Eq. (\ref{fitting}).
}
\label{figure3}
\end{figure}

We first introduce the edges along the $y$ axis in two-dimensional torus geometry by replacing $\delta_{i,j} '$ with the usual Kronecker delta $\delta_{i,j}$, which prohibits electron transfer to different unit cells and introduces translational symmetry breaking in the FQH states. To study the edge state in the FQH system, we adjust the number of electrons to be $\nu=1/3$ in the bulk. Figure \ref{figure2} shows the occupation number of the  one-particle state $\varphi_j({\bm r})$ for various sizes of systems with a total number of electrons $N_e=M/3$ and an aspect ratio $L_y/L_x$ of the unit cell. The guiding center $X_j$ corresponds to the $x$ coordinate of the center position of $\varphi_j({\bm r})$. Note that we plot the expectation value of the number operator $\langle n_j \rangle = \langle c^\dagger_j c_j \rangle$ instead of the charge density itself. 
The latter is obtained by using the wave function $\varphi_j({\bm r})$, which smears the detailed structure of the edge states. In Fig.~2, we confirm the uniform electron density of the $\nu=1/3$ FQH state characterized by the Laughlin state. In the region near the edges, however, we find oscillations of electron density. These features are almost independent of both the aspect ratio $L_y/L_x$ and the size of the unit cell, and the oscillations are well fitted by 
\begin{align}\label{fitting}
f_\nu (x) = C_\nu {\rm exp}(-x/\xi_\nu ){\rm cos}(k_\nu x+\theta_\nu )+\nu,
\end{align}
where $C_\nu, \xi_\nu, k_\nu$, and $\theta_\nu$ are fitting parameters corresponding to the filling factor $\nu$. 
The inset in Fig.~2 shows the fitting result for $N_e=40, L_y/L_x=0.5$, and the 
optimized values of $k_{1/3}$ and $\xi_{1/3}$ are $1.46 l^{-1}$ and $2.79l$, respectively. The wave number $k_{1/3}$ is in good agreement with the wave number of the bulk magnetoroton minimum\cite{Yoshioka1986}, which means the density oscillation is characterized by low-energy collective excitations induced by the boundary conditions at the edge. The presence of edge roton excitation was reported by several previous works\cite{Hawrylak1996,Yanagi2005,Jolad2010}. To find the general feature of the edge states, we additionally investigate the $\nu=2/5$ FQH state. The obtained result for $N_e=56, L_y/L_x=0.5$, and $M=140$ is shown in Fig.~3, which indicates a similar fitting by Eq. (5) reproduces the density oscillation induced by the boundary condition. The fitting parameters are obtained as $k_{2/5}=1.68l^{-1}$ and $\xi_{2/5}=3.90l$, reflecting a higher electron density and a smaller bulk excitation gap of the FQH state. The obtained ratio $\xi_{2/5}^{-1}/\xi_{1/3}^{-1}=0.71$ is close to the previously estimated gap ratio of roton excitation of $\sim 0.67$\cite{Morf2002}, which also confirms that the edge structure is related to the bulk collective charge excitations.

Besides our numerical analysis, the edge density profile of the FQH states is
studied in connection with the Hall viscosity\cite{Park2014,Zhu2020}.
It is interesting that the microscopic edge structure is also related to the
topological properties.

\subsection{Edge reconstructions by confinement potential}

\begin{figure}[t]
\centering
\includegraphics[scale=0.7]{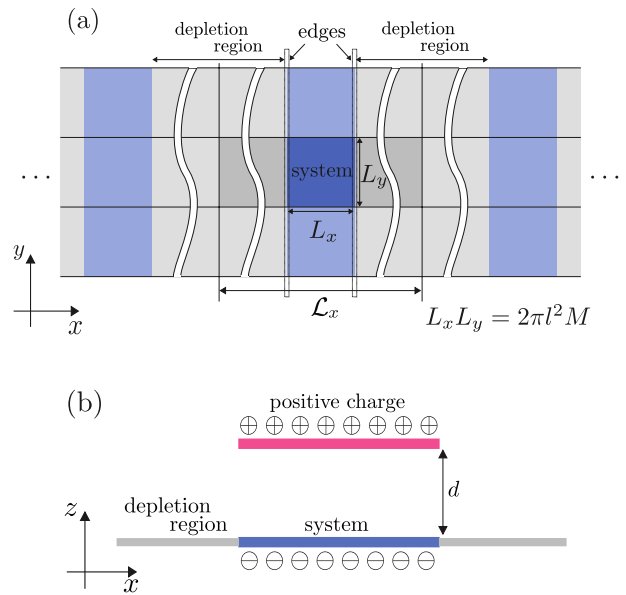}
\caption{(a) The extended unit cell used in the calculation.
${\cal L}_x$ is the extended periodicity in the $x$ direction. The depletion region 
is illustrated in gray where electrons are absent. When ${\cal L}_x$ is sufficiently large, the Coulomb interaction from the other unit cells across the edge is safely omitted. (b) The model of the confinement potential originated from the uniform positive background charges. Following Wan \textit{et al.}\cite{Wan2002}, we suppose 
a layer of uniform positive background charge at vertical distance $d$. }
\label{figure4}
\end{figure}

\begin{figure}[t]
\centering
\includegraphics[scale=0.35]{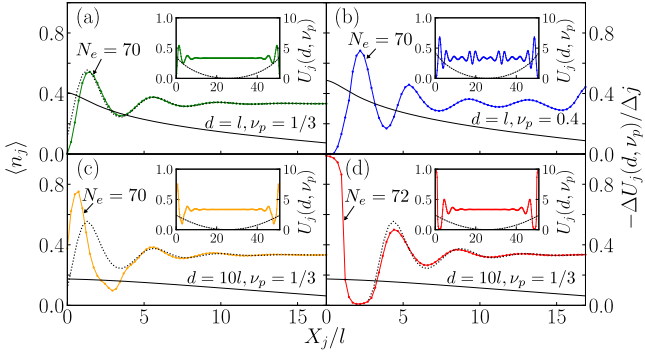}
\caption{Density oscillations induced by the confinement potential given by Eq.(6). 
(a) Roton-minimum-type edge structure for small $d$ in the charge neutral condition $\nu_p=1/3$. 
(b) Bulk excitations caused by $\nu_p > 1/3$ for small $d$. 
(c) Modified edge structure for large $d$. 
(d) Reconstructed edge structure composed of $\nu=1$ and $\nu=1/3$ edge states
for a slightly increased number of electrons from $N_e=M/3$. 
The black dotted line shows the result for $N_e = 40, L_y/L_x=0.5$ in Fig.~2.
The black solid line indicates the differential of the confinement potential $-\Delta U_j(d,\nu_p)/ \Delta j$, 
and the black dashed line in the insets indicates the confinement potential $U_j(d,\nu_p)$.}
\label{figure5}
\end{figure}

The system used in the above calculations does not include the effect of Coulomb potential 
from the positive ions near the conduction electrons. 
Instead, the Coulomb potential from the electrons in neighboring unit cells is used as an effective 
confinement potential, which is self-consistently optimized to reduce the Coulomb energy and expected to realize
the fundamental edge structure caused by the breaking of the translational symmetry.
In realistic systems, however, the confinement potential originates mainly from the positively charged background ions near the conduction electrons and the metal gates placed on the sample. To investigate the edge states of the FQH systems in a more realistic situation, we extend the system as follows: First, we expand the length $L_x$ of the unit cell to ${\cal L}_x$ and attached depletion region where the electrons are absent, as shown in Fig.~4(a). When the depletion region is sufficiently large, the effect from the electrons in different unit cells across the edges is safely neglected. We estimated the Hartree-Fock potential from other unit cells and confirmed that ${\cal L}_x/L_x \geq 21$ is sufficient to omit its position dependence. Second, following Wan {\it et al.}\cite{Wan2002}, we introduce the uniform positively charged layer at the vertical distance $d$ from the electron layer, as shown in Fig.~4(b).
The Coulomb potential for the 2D electrons is then given as
\begin{align}
U_j(d,\nu_p) &= \int_{0}^{{\cal L}_x}dx_e \int_{0}^{L_y}dy_e \int_{0}^{{\cal L}_x}dx_p \int_{0}^{L_y}dy_p  \times \notag \\
&{\cal U}(\bm{r}_e,\bm{r}_p,d,\nu_p)\varphi_j^*(\bm{r}_e)\varphi_{j}(\bm{r}_e),
\end{align}
where
\begin{align}
&{\cal U}(\bm{r}_e,\bm{r}_p,d,\nu_p) \notag \\ 
&= \sum_{\substack{n_x,n_y,\\m_x,m_y \in \mathbb{Z}}}\frac{-e^2}{\varepsilon} \frac{\sigma({\bm r}_p+m_x{\cal L}_x{\bm e}_x+m_yL_y{\bm e}_y,\nu_p)}{|({\bm r}_e-{\bm r}_p)+n_x{\cal L}_x{\bm e}_x+n_yL_y{\bm e}_y + d {\bm e}_z|},
\end{align}
\begin{align}
\sigma({\bm r},\nu_p) &= 
\begin{cases}
\frac{\nu_p}{2\pi l^2} & (\frac{n{\cal L}_x}{2} -\frac{L_x}{2} \le x \le \frac{n{\cal L}_x}{2} +\frac{L_x}{2}, n \in \mathbb{Z})\\
0 & \mbox{otherwise},
\end{cases}
\end{align}
and ${\bm r}_e$ and ${\bm r}_p$ are coordinates of electrons and positive ions within the unit cell, respectively. $\nu_p$ is the effective filling factor of positive ions relative to the number of magnetic fluxes. Although the 
densities of dopant and conduction electrons are balanced in the usual situation (\textit{i.e.} $\nu=\nu_p$), we use $\nu_p$ as a variable parameter to study the effects of gate voltage. 
Hereafter, we set $M=210, L_y/L_x=0.5$, and ${\cal L}_x/L_x=21$.

To confirm the consistency with the results obtained in the previous section,
we first set $d=l$ and $\nu_p=\nu =1/3$. Since an almost uniform potential is obtained in the region of small $d$ under the charge neutral condition $\nu = \nu_p$, a structure similar to the previous results shown in Fig.~2 is expected. As presented in Fig.~5(a), the obtained result (green line) is in good agreement
with the previous one (black dotted line). 

When we increase the density of positive ions $\nu_p$, the confinement potential is enhanced, and the electron density is modified to reduce the total potential energy. 
As seen in Fig.~5(b), the differential of the confinement potential (black solid line) for $\nu_p = 0.4$ is enhanced near the edge, and part 
of the electrons move to the bulk region, yielding charge excitations in the bulk. 
To check the elementary charge of this excitation, we additionally calculate the accumulated deviation $D_j$, which is defined as
\begin{align}
 D_j = \sum_{i=0} ^j (\langle n_i \rangle - 1/3).
\end{align}
Since the total electron filling is set as $\nu=1/3$, the jump in $D_j$ is interpreted as local charge accumulation. As shown in Fig.~6, $D_j$ shows a clear jump by passing the bulk excitations, and the amount of $\Delta_n = 1/3$ means the excitations are fractional quasiparticles.
This result implies that depending on the strength of the potential, bulk quasiparticle excitations are created. This nature may be related to recent experimental bulk transport at the $\nu=1/3$ FQH state\cite{Altimiras2012,Inoue2014}.

\begin{figure}[t]
\centering
\includegraphics[scale=0.6]{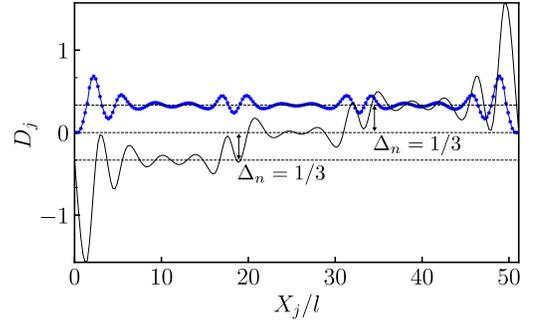}
\caption{The black solid line shows the accumulated deviation 
from the uniform density of $\nu=1/3$ defined by $\sum_{i=0} ^j (\langle n_i \rangle - 1/3)$.
The clear jumps of $\Delta_n=1/3$ indicate the bulk excitations found in Fig.~5(b)
are fractional quasiparticles of the $\nu=1/3$ Laughlin state. 
The blue line shows the original density of electrons $\langle n_i \rangle$ in Fig.~5(b).
}
\label{figure6}
\end{figure}

To see the potential form dependence of the edge state, we next modify the vertical distance $d$ of positive ions. Figure 5(c) shows the result at $\nu_p=1/3$ and $d=10l$, which is a typical value of realistic samples. Since the large distance $d$ weakens the confinement force near the edge (see the solid black line), the Coulomb repulsion between the electrons dominates at the edge, and the electrons split into two domains. This result is consistent with the previous work by Wan \textit{et al.}\cite{Wan2002} and is called edge reconstruction. Comparing the edge roton-type density oscillation plotted as a dotted line, we find the difference appears only in the outermost edge region, which implies the edge reconstruction occurs independently of the bulk FQH states. 
More clear domain splitting is observed at $N_e=72$, where the electrons are slightly 
increased from $\nu=1/3$, as shown in Fig.~5(d).
We find a clear domain of $\nu=1$ at the outermost region. Hence, additional $\nu=1$ counterpropagating edge modes are expected to appear, and roton-minimum-type density oscillation is reproduced as the inner structure. This result is understood as follows: Since the Coulomb interaction dominates at the edge under smooth confinement potential, the electrons added to the system are repelled by the electrons in the central bulk region and accumulate at the boundary of the sample. Once the domain of $\nu=1$ is formed, the electrons in the outermost domain behave as static charges and work as an effective confinement potential which stabilizes the inner edge structure. As is shown later, the inner edge structure is stable even if the size of the outermost $\nu=1$ domain changes, which suggests this pair of $\nu=1$ and $\nu=1/3$ edge structures reduces both the potential energy from the positive ions and the Coulomb energy between the electrons.

\begin{figure}[t]
\centering
\includegraphics[scale=0.38]{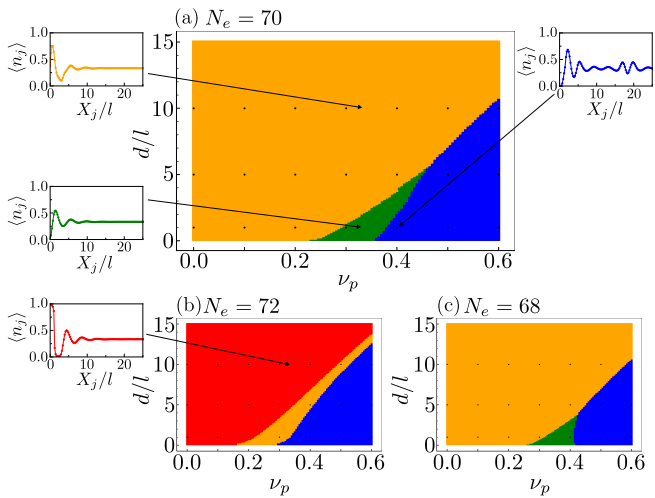}
\caption{Phase diagram of the edge states for various $d$ and $\nu_p$ at (a) $N_e=70$, (b) $N_e=72$, and (c) $N_e=68$. The colors of the regions correspond to those of structures shown in Fig.~5.
The black dots show the parameters used in the determination of the electron densities.
The system size of $M=210$ is fixed.}
\label{figure7}
\end{figure}

We next describe the phase diagram of the edge structure by categorizing the above response to the confinement potential shape.
For this purpose, we move $\nu_p$ from $0$ to $0.6$ in $0.1$ increments at $d/l=1,5,10$ and determine the phase boundaries for three different total numbers of electrons, $N_e= 68, 70$, and $72$ at $M=210$.
The results are presented in Fig.~7. We find the roton-minimum-type edge structure in the range of $\nu_p=0.25$--$0.45$ and $d<5l$ for $N_e=68$ and $70$. This behaviour is in good agreement with the previous work reporting the edge reconstructions for $d$ exceeding $1.5l$ at $\nu_p=1/3$\cite{Wan2002,Wan2003}. 
Our result indicates that the edge structures are roughly classified into two groups: the roton-minimum-type edge structure with or without bulk excitations (blue and green regions), and the reconstructed edge structures (red and orange regions). We note that the conventional roton-minimum-type edge structure is unstable in the absence of positive charge ions, $\nu_p=0$, 
while it is stabilized by the presence of the $\nu=1$ outermost domain.

Finally, we investigate the response to the shift of the chemical potential. 
We choose the parameters to be $d=10l, \nu_p=1/3$ [the same as in Fig.~5(c)] and  $d=l, \nu_p=1/3$ [the same as in Fig.~5(a)] to see how the previous results are modified with the change in the total number of electrons $N_e$.
Figure 8(a) shows the results when we increase $N_e$ from $70$ at $d=10l$. 
We find a clear outermost $\nu=1$ domain, which works as an absorber and retains the inner roton-minimum-type edge structure. On the contrary, 
as shown in Fig.~8(b), the decrease in $N_e$ modifies the edge structure 
complicatedly.
Since a weak confinement potential of $d=10l$ effectively enhances the repulsive 
interaction between the electrons, the edge structure deforms to 
reduce the electron density, and various electron configurations
appear at the edge. 
When we again increase $N_e$ from 70 at $d=l$,
the additional electrons are absorbed in the outermost peak, and the inner structure deforms slightly. This structure is a transient structure to that shown in Fig.~8(a). When we decrease $N_e$, the roton-minimum-type edge structure remains to some extent, as shown Fig.~8(d). 
Although the decrease of $N_e$ causes deformation of the edge structure similar to that in the case of $d=10l$, the edge structure at $N_e=68$ is exactly the same as that of $N_e=70$, which indicates that under such a potential form, the roton-minimum-type edge structure is robust against chemical potential variation. Figure 8 also shows that the electron density in the bulk region is not affected by the change in the number of electrons, which is clear evidence of the edge compressibility and bulk incompressibility.

\begin{figure}[t]
\centering
\includegraphics[scale=0.35]{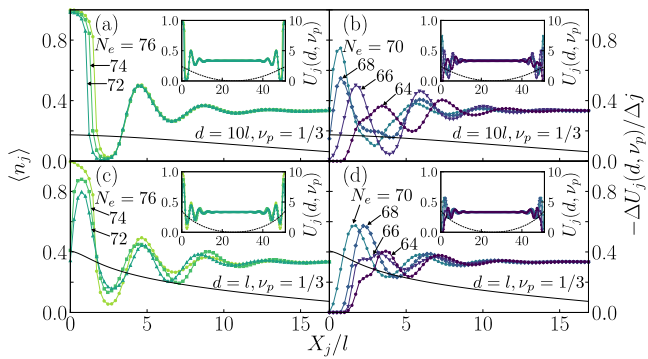}
\caption{Chemical potential dependence of the edge structure at $\nu_p=1/3$.
$d=10l$ for (a) and (b), and $d=l$ for (c) and (d). 
The total number of electrons varies in the range of $N_e = 76$--$64$ under the fixed system size of $M=210$.
The black solid line indicates the differential of the confinement potential $-\Delta U_j(d,\nu_p)/ \Delta j$, 
and the black dashed line in the insets indicates the confinement potential $U_j(d,\nu_p)$.}
\label{figure8}
\end{figure}

\section{Summary}

In this paper, we have studied the FQH edge structure at filling factor $\nu=1/3$ using the DMRG method. The obtained density oscillation near the edge is explained by locally excited magnetorotons, indicating the edge state is characterized by the bulk properties of the FQH state. We have also investigated the potential form and chemical potential dependencies by introducing a realistic confinement potential taking into account positive background charges. The increase in the positive charges from the charge neutral condition enhances the confinement potential and causes fractional charge excitations in the bulk region, while the increase in vertical distance to the positive background charge weakens the confinement and induces edge reconstruction. The chemical potential dependence indicates the appearance of a $\nu=1$ domain that stabilizes the inner $\nu=1/3$ edge structure, which suggests the edge reconstruction occurs in the process of reducing both the potential energy and the Coulomb energy of electrons. The condition-independent bulk state that is stable against the change in the number of the electrons also shows that the edge state is compressible, while the bulk part is incompressible. 

\begin{acknowledgments}
The authors acknowledge K. Muraki and M. Hashisaka for valuable discussions. 
This work was supported by JSPS KAKENHI Grant No.~JP19K03708.

\end{acknowledgments}

\bibliographystyle{apsrev4-1}
%\bibliography{reference.bib}
%merlin.mbs apsrev4-1.bst 2010-07-25 4.21a (PWD, AO, DPC) hacked
%Control: key (0)
%Control: author (72) initials jnrlst
%Control: editor formatted (1) identically to author
%Control: production of article title (-1) disabled
%Control: page (0) single
%Control: year (1) truncated
%Control: production of eprint (0) enabled
%
\end{document}